\documentclass[aps,prl,twocolumn,superscriptaddress]{revtex4-1}

\usepackage{graphicx}
\usepackage{subfigure}
\usepackage{natbib}
\usepackage{amsmath,amssymb,amsfonts} 
\usepackage[usenames,dvipsnames]{color}
\begin{document}

\title{Laser-produced magnetic-Rayleigh-Taylor unstable plasma slabs\\in a 20~T magnetic field}

\author{B.~Khiar}\affiliation{Sorbonne Universit\'e, Observatoire de Paris, PSL Research University, LERMA, CNRS UMR 8112, F-75005, Paris, France}\affiliation{Flash Center for Computational Science, University of Chicago, 5640 S. Ellis Avenue, Chicago, IL 60637, USA}

\author{G.~Revet}\affiliation{LULI - CNRS, CEA,   Sorbonne Universit\'e, Ecole Polytechnique, Institut Polytechnique de Paris - F-91128 Palaiseau cedex, France}\affiliation{Institute of Applied Physics, RAS, 46 Ulyanov Street, 603950 Nizhny Novgorod, Russia}

\author{A.~Ciardi}\affiliation{Sorbonne Universit\'e, Observatoire de Paris, PSL Research University, LERMA, CNRS UMR 8112, F-75005, Paris, France}

\author{K.~Burdonov}\affiliation{Sorbonne Universit\'e, Observatoire de Paris, PSL Research University, LERMA, CNRS UMR 8112, F-75005, Paris, France}\affiliation{LULI - CNRS, CEA,   Sorbonne Universit\'e, Ecole Polytechnique, Institut Polytechnique de Paris - F-91128 Palaiseau cedex, France}\affiliation{Institute of Applied Physics, RAS, 46 Ulyanov Street, 603950 Nizhny Novgorod, Russia}

\author{E.~Filippov}\affiliation{Institute of Applied Physics, RAS, 46 Ulyanov Street, 603950 Nizhny Novgorod, Russia}\affiliation{Joint Institute for High Temperatures, RAS, 125412, Moscow, Russia}

\author{J.~B\'eard}\affiliation{LNCMI, UPR 3228, CNRS-UGA-UPS-INSA, 31400 Toulouse, France}

\author{M.~Cerchez}\affiliation{Institute for Laser and Plasma Physics, University of D\"usseldorf, 40225 D\"usseldorf, Germany}

\author{S.N.~Chen}\affiliation{ELI-NP, "Horia Hulubei" National Institute for Physics and Nuclear Engineering, 30 Reactorului Street, RO-077125, Bucharest-Magurele, Romania}

\author{T.~Gangolf}\affiliation{LULI - CNRS, CEA,   Sorbonne Universit\'e, Ecole Polytechnique, Institut Polytechnique de Paris - F-91128 Palaiseau cedex, France}\affiliation{Institute for Laser and Plasma Physics, University of D\"usseldorf, 40225 D\"usseldorf, Germany}

\author{S.S.~Makarov}\affiliation{Joint Institute for High Temperatures, RAS, 125412, Moscow, Russia}

\author{M.~Ouill\'e}\affiliation{LULI - CNRS, CEA,   Sorbonne Universit\'e, Ecole Polytechnique, Institut Polytechnique de Paris - F-91128 Palaiseau cedex, France}

\author{M.~Safronova}\affiliation{LULI - CNRS, CEA,   Sorbonne Universit\'e, Ecole Polytechnique, Institut Polytechnique de Paris - F-91128 Palaiseau cedex, France}\affiliation{Institute of Applied Physics, RAS, 46 Ulyanov Street, 603950 Nizhny Novgorod, Russia}

\author{I.Yu.~Skobelev}\affiliation{Joint Institute for High Temperatures, RAS, 125412, Moscow, Russia}\affiliation{National Research Nuclear University MEPhI, 115409 Moscow, Russia}

\author{A.~Soloviev}\affiliation{Institute of Applied Physics, RAS, 46 Ulyanov Street, 603950 Nizhny Novgorod, Russia}

\author{M.~Starodubtsev}\affiliation{Institute of Applied Physics, RAS, 46 Ulyanov Street, 603950 Nizhny Novgorod, Russia}

\author{O.~Willi}\affiliation{Institute for Laser and Plasma Physics, University of D\"usseldorf, 40225 D\"usseldorf, Germany}

\author{S.~Pikuz}\affiliation{Joint Institute for High Temperatures, RAS, 125412, Moscow, Russia}\affiliation{National Research Nuclear University MEPhI, 115409 Moscow, Russia}

\author{J.~Fuchs}\affiliation{LULI - CNRS, CEA,   Sorbonne Universit\'e, Ecole Polytechnique, Institut Polytechnique de Paris - F-91128 Palaiseau cedex, France}\affiliation{Institute of Applied Physics, RAS, 46 Ulyanov Street, 603950 Nizhny Novgorod, Russia}

\begin{abstract} 
Magnetized laser-produced plasmas are central to many novel laboratory astrophysics and inertial confinement fusion studies, as well as in industrial applications. Here we provide the first complete description of the three-dimensional dynamics of a laser-driven plasma plume expanding in a 20~T transverse magnetic field. The plasma is collimated by the magnetic field into a slender, rapidly elongating slab, whose plasma-vacuum interface is unstable to the growth of the ``classical'', fluid-like magnetized Rayleigh-Taylor instability.

\end{abstract}
\maketitle

The combination of high-power lasers with externally applied high-strength magnetic fields of up to kT \cite{fujioka_kilotesla_2013,PhysRevE.95.053204} has been seminal in the development of many recent applications in laboratory astrophysics\cite{Revete1700982, Ciardi2013,Albertazzi325, schaeffer_high-mach_2017, kuramitsu_magnetic_2018}, in novel concepts in laser-\cite{Perkins2013,Montgomery2015,Davies2017} and magnetically-driven\cite{Slutz2010,Slutz2012} inertial confinement fusion physics, and in industrial applications\cite{Haverkamp2009,Ho2014}. Beside understanding the dynamics of the plasma expansion across a magnetic field, of particular importance is to grasp the nature of rapidly growing instabilities which may develop and profoundly modify the morphology and characteristics of these plasmas. Indeed, the presence of striations and flutes have often been associated with the development of instabilities and in particular with the lower hybrid drift instability (LHDI) or one of its variants\cite{PhysRevA.4.2094, gladd_lower_1976, JGRA:JGRA8826}. In addition, anomalous resitivity driven by the LHDI\cite{davidson_anomalous_1977,choueiri_anomalous_1999} can also affect the plasma microscopically, with potentially important consequences on magnetic field diffusion and the growth of other instabilities. Among those, the magnetic Rayleigh-Taylor instability (MRTI)\cite{Kruskal348,chandrasekhar1961hydrodynamic} is known to play a key role on the dynamics of laboratory\cite{RyutovZpinch2000}, as well as astrophysical plasmas\cite{isobe_filamentary_2005,hester_1996}. So far however, it has not been isolated in laser-produced high energy density plasmas.
    
A major parameter affecting the stability and dynamics of these plasmas is the relative direction of the applied magnetic field with respect to the plasma expansion axis. While for an \textit{aligned} magnetic field the plasma is collimated into an axisymmetric, stable jet-like flow \cite{Ciardi2013,Albertazzi325}, for a \textit{transverse} magnetic field both stable\cite{PhysRevLett.111.185002} and unstable flows\cite{PhysRevLett.62.2837} were observed and a clear understanding of the plasma evolution is still missing.

Here, we provide the first complete description of the three-dimensional dynamics of a laser-driven plasma plume in a \textit{transverse} 20 T magnetic field. We show that the plasma is collimated into a slender, rapidly expanding slab, and demonstrate that under these conditions, the growth of flute-like, interchange modes at the plasma-vacuum interface that extend in the form of spikes into the vacuum is due to the classical, fluid-like, magnetic Rayleigh-Taylor instability (MRTI). Interestingly, we find that to recover quantitatively in the simulations the penetration of these spikes into the vacuum, a subgrid-scale model of anomalous resistivity needs to be included. This anomalous resistivity could be induced by the micro-turbulence generated by the LHDI, which for our plasma conditions grows over very fast time scales and short spatial scales.

\begin{figure}
\includegraphics[width=8.6cm]{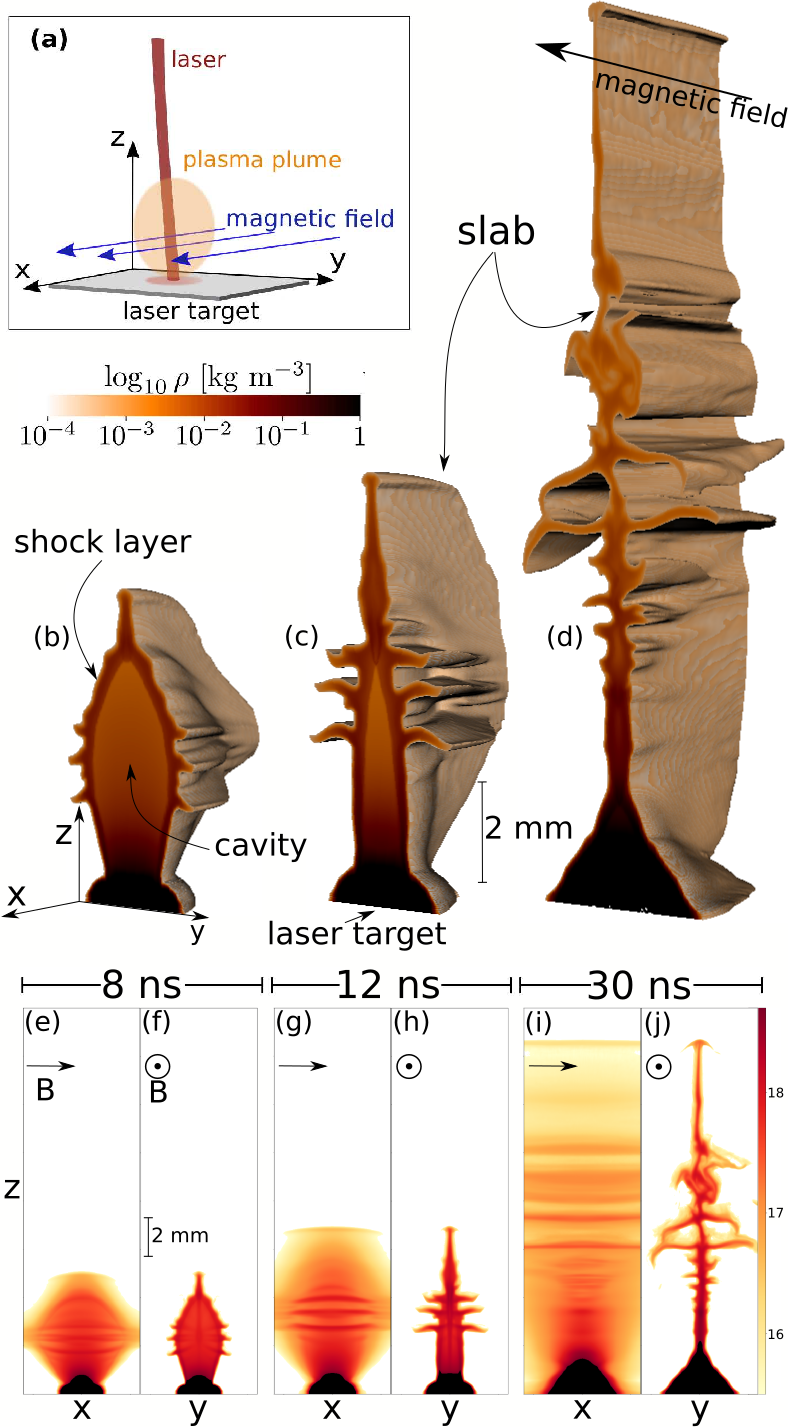}
\caption{Schematic of the experimental set up (a). The external uniform magnetic field of 20 T is initially oriented along the $x$-axis. Panels (b), (c) and (d) show the simulated, three dimensional mass density distribution at different times (8, 20 and 48 ns). To show the density distribution inside the flow and the various structures discussed in the text, only the $x<0$ part of the computational domain is rendered. Panels (e-j) show the decimal logarithm of the electron number density (cm$^{-2}$) integrated \textit{along} or \textit{perpendicularly} to the magnetic field.\label{fig3D}}
\end{figure}

The experiments are performed on three different facilities, namely ELFIE, TITAN and PEARL, with similar laser parameters (see Supplemental Material, which also includes Refs. \cite{Vinci2018N,Higginson2015,faenov_1994,FLYCHK,ryazantsev_2016,filippov_parameters_2016}). In all cases, the on-target intensity of the nanosecond-duration laser pulse is kept the same, $I \sim 10^{13}$ W cm$^{-2}$. The laser irradiates a Teflon, (C$_2$F$_4$)$_n$, foil target placed in a vacuum in the presence of an externally applied, pulsed ($\sim \mu$s) magnetic field \cite{Higginson2017}. The magnetic field is initially parallel to the target surface (Fig. \ref{fig3D}a). Note that we do not observe any significant modification of the overall plasma dynamics despite using different focal spot sizes in the different experiments. The magnetic field was created in each experiment using a pulsed-power driven Helmholtz coil (see Supplemental Material). The main difference comes from the field strength, which is of 20 T on ELFIE and TITAN, while it is limited to 13.5 T on PEARL. Under our conditions, the magnetic field generated via the Biermann battery, which is active only while the laser irradiation is maintained and limited to low strengths, $\simeq$ 1 T, beyond 1 mm of expansion \cite{lancia,gao2015}, has negligible dynamical effects on the plasma, especially considering the large spatial and temporal scales investigated here. As shown experimentally in  \cite{Higginson2017} the presence of an applied field is crucial in collimating the plasma plume.
    
The experimental data are complemented by three-dimensional, single-fluid, bi-temperature resistive magneto-hydrodynamic (MHD) simulations carried out with the code GORGON \cite{Chittenden2004,ciardi2007}. The model includes a correction to the resistivity due to lower-hybrid drift microinstabilities \cite{Chittenden1995}. The computational vacuum cut-off density is set to $10^{-4}$~kg m$^{-3}$ and the simulated domain $(L_x,L_y,L_z)$ extends over ($8 \times 6 \times 20) $~mm with a spatial resolution of $20$~$\mu$m. As in previous work\cite{Ciardi2013,Albertazzi325,Revete1700982}, The initial laser interaction with a solid Carbon target was modeled using the Lagrangian, radiation hydrodynamic code DUED\cite{Atzeni2005} in 2D axisymmetric, cylindrical geometry, and then remapped onto the GORGON grid. 

\begin{figure}
\includegraphics[width=7.5cm]{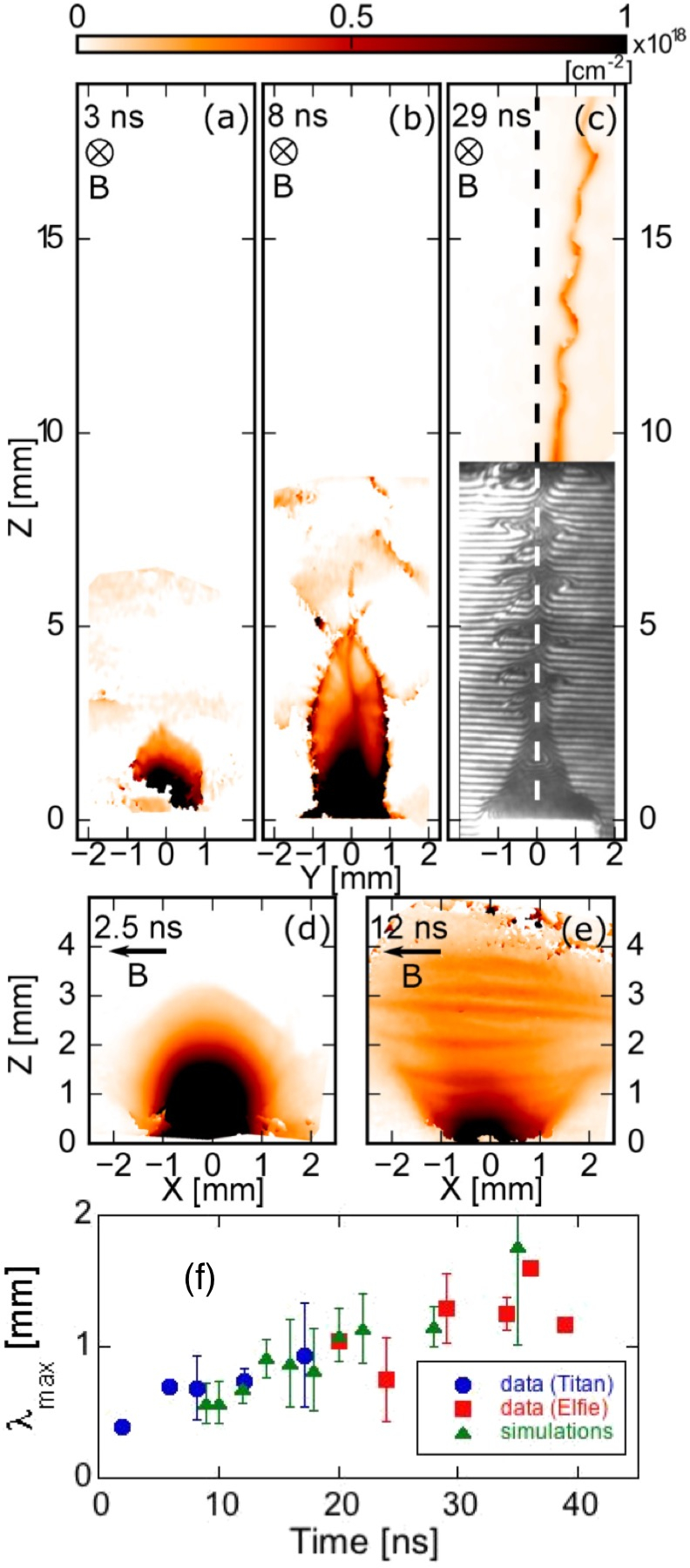}
 \caption{Maps of integrated electron areal density ($\int n_e dl~[cm^{-2}]$) probed parallel (a,b, and c, ELFIE data) and perpendicular to the magnetic field  (d and e, TITAN data), at different times. The density is deduced from the fringe shift of the raw interferograms with respect to a reference image with no plasma. In the lower part of panel (c) we show the raw fringe image, the integrated density is too large and the probe beam light is lost in the transit through the dense part of the plasma slab (seen as the dark regions where fringes are absent). Nevertheless, the raw image clearly shows the MRTI spikes pattern. Panel (f) shows the variation of the spatial separation between the large-scale spikes as a function of time. As indicated in the legend, data at early times are inferred from TITAN shots,  while data at late times are inferred from ELFIE shots. Overlaid are measurements from the GORGON simulations (see Supplemental Material for details).  \label{fig_experiment}}
\end{figure}

The overall three-dimensional plasma dynamics and the development of the instabilities is presented in Fig. \ref{fig3D} and \ref{fig_experiment} for the simulation and experiments respectively. We show experimental data from ELFIE (Fig. \ref{fig_experiment}.a-c with probing parallel to the magnetic field) and TITAN (Fig. \ref{fig_experiment}.d-e with probing perpendicular to the magnetic field) experiments. On PEARL, probing was performed simultaneously along the two directions (see Supplemental Material), confirming the global, thin slab-like development of the plasma seen at ELFIE and TITAN, except that the instability was less developed due to the lower field magnetic field strength. The initial plasma expansion ($\lesssim 3$ ns) is unconstrained by the magnetic field, and it is characterized by a very large dynamic plasma-$\beta$, $\beta_{dyn}=2 \mu_0 \rho v^2 / B^2  \sim 10^{3}$. Because of the relatively large electron temperatures in the plume ($T_e\sim 100\--300$~eV)\cite{Higginson2017}, the plasma is highly conductive and the magnetic field is advected with the flow. This  ideal magnetohydrodynamic regime is characterized by a relatively large magnetic Reynolds number,  $\text{Re}_{m}=Lv/D_m\sim 100$, where $L\sim 10^{-3}$~m, $v \sim 10^5$~m.s$^{-1}$ and $D_m \sim 1$ m$^2$s$^{-1}$ are respectively the characteristic length, velocity, and magnetic diffusivity. Furthermore, both thermal conduction and viscosity are unimportant in the initial formation of the cavity and slab (see labels in Fig. \ref{fig3D}b and \ref{fig3D}c), Peclet and Reynolds numbers, $\text{Pe} \sim10$ and $\text{Re} \sim 10^4$. Plasma expansion, which occurs at speeds $2-3$ times larger than the fast magneto-acoustic speed $c_{ma}=\sqrt{c_{s}^{2}+v_{A}^{2}}$ , leads to the compression ($B_{\max}\sim 27$~T) and bending of the magnetic field lines at the edge of the plasma  ($c_{s}$ and  $v_{A}$ are the sound and Alfven speeds respectively).  The ensuing deceleration gives rise to a reverse shock in the expanding flow and the formation of a shell of shocked plasma with a width $\delta_{sl}\sim 200\,\mu m$. As we shall discuss later, it is at the interface between this shocked plasma and the vacuum that the MRTI develops. The presence of a lower density cavity delimited by an envelope of shocked plasma is clearly seen in the experimental and simulation data after a few nanoseconds of expansion (Fig. \ref{fig_experiment}b and Fig. \ref{fig3D}b).  However, we observe that the flow becomes later on highly asymmetrically in the directions parallel and perpendicular to the initial magnetic field. In the $y-z$ plane the generation of a jet-like flow is similar to the case where the magnetic field is aligned with the main expansion axis of the flow (i.e. perpendicular to the target)\cite{Ciardi2013, Albertazzi325,Higginson2017}. It is the result of the curved shock layer re-directing the plasma flow towards the tip of the cavity, where a conical shock re-collimates the flow in the $z$-direction. However, in the $x-z$ plane, the plasma flow is unconstrained by the magnetic field, and as a result, the flow takes the shape of a thin ($\delta y \sim 0.4 - 0.8 $~mm) magnetically confined plasma "pancake". The experiments indicate that by $\sim 30$~ns the plasma has reached a distance $z\sim 20$~mm (Fig. \ref{fig_experiment}c), corresponding to propagation speeds $\sim 600$ km s$^{-1}$, in agreement with the simulations.

Alongside the general features presented so far, we find clear evidence of the rapid (few ns) growth of flute-like filaments with a characteristic wavelength perpendicular to the magnetic field $\sim 1$ mm (see Fig. \ref{fig3D}c,g,h, Fig. \ref{fig_experiment}c,e, and Fig. \ref{fig_experiment}f). In general, the plasma flow undergoes a succession of lateral expansion and contractions, during which further filaments may be generated. These filaments are well aligned with the magnetic field and protrude into the vacuum to a distance $|y|\sim 1$~mm (Fig. \ref{fig3D}c, h and Fig. \ref{fig_experiment}c). Similar flute-like structures were observed in laser experiments with similar intensities, $I\sim 10^8\--10^{13}$~W~cm$^{-2}$ but lower magnetic field, $B\sim 0.01\--1.5$~T, and the cause was tentatively attributed to the LHDI or one of its variants\cite{Okada1981,ripin_1987,Dimonte1991}. These instabilities are driven by cross-field currents in an inhomogeneous plasma and magnetic field, where the electrons are magnetized and the ions are not. In our experiments, the LHDI growth rate is approximately the lower hybrid frequency, $\gamma\lesssim w_{LH}\sim 10^{11}$~s$^{-1}$, and the dominant wavelengths is roughly the electron gyroradius, $\lambda \sim 2 \pi (T_i/2T_e)^{1/2} r_{L,e}\sim 14$~$\mu$m; the numerical values quoted are for our nominal plasma parameters $B\sim 20$~T, $n_i\sim 10^{18}$~cm$^{-3}$, $T_i\sim 500$~eV, $T_e\sim200$~eV and $\left\langle Z \right\rangle\sim 6$.  Moreover, in our case the electrons (and ions) are collisional,  $\tau_e=\nu_e^{-1}\sim 10^{-11}$~s, and only perturbation with wavelengths $\lambda\lesssim 2\pi \left(v_{T,i}/\nu_{e}\right)\left(r_{L,i}/L_{n}\right)^{2}\sim 3$~$\mu$m are expected to grow\cite{1981JATP...43..775H}; where $L_n\sim 100$~$\mu$m  is the density gradient scale-length and $v_{T,i}=(2k_BT_i/m_i)^{1/2}$ is the ion thermal velocity. From these estimates it is clear that the time and length scales associated with the LHDI are orders of magnitude smaller than the growth times and the wavelengths of the density filaments we observe ($\gamma^{-1}\sim2$~ns and $\lambda\gtrsim 500$~$\mu$m, c.f. Fig. \ref{fig3D}c,e-j and Fig. \ref{fig_experiment}c,e,f). However the micro-turbulence generated by the LHDI can still readily enhance the electrical resistivity of the plasma\cite{choueiri_anomalous_1999}. This is similar to simulations of z- or $\theta$-pinches, where an accurate modeling of the highly-magnetized but low-density (nearly vacuum) plasma regions requires the inclusion of an anomalous resistivity\cite{Chittenden1995}. We find that in our simulations this is particularly important to allow the spikes to expand into the vacuum to distances, $\sim 1$~mm, consistent with the experimental observations (Fig. \ref{fig_experiment}c). However, we stress that the inclusion of an anomalous resistivity in the simulations does not alter the characteristic growth time-scales or wavelengths of the dominant MRTI modes. These develop in regions of relatively higher plasma densities where anomalous resistive effects are negligible.

\begin{figure}
\includegraphics[width=8.6cm]{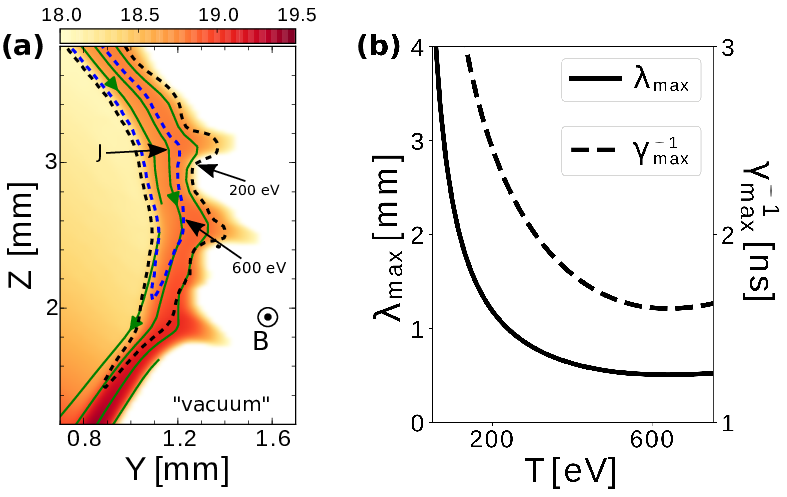}
\caption{Estimates of the MRTI growth time and the fastest growing mode. (a) Zoom of the plasma/vacuum interface at $t=8$~ns. The color map corresponds to $\log_{10} n_e$ in cm$^{-3}$. Green lines show contours of the magnitude of the current density. The dashed line contours show the ion temperature. Panel (b) shows the temperature dependence of the fastest growing mode for the MRTI instability in the presence of resistivity and viscosity. At low temperatures resistive damping dominates whereas at high temperatures the viscous dissipation is predominant. The curve is for a fully ionized carbon plasma of density $\rho=0.02$~kg m$^{-3}$ and with $\Lambda = 9$. \label{figInstability}}
\end{figure}
 
Contrary to previous work carried out at lower magnetic field strengths \cite{PhysRevLett.62.2837,Dimonte1991}, the growth of large-scale spikes seen here for $z<10$~mm is consistent with the classical, fluid-like, MRTI. This instability can grow at a plasma-vacuum interface, or at an interface separating different density mediums, when the effective acceleration in the frame of the interface is anti-parallel to the density gradient. In our case the density gradient always points towards the interior of the cavity, while the effective acceleration on the plasma envelope due to Lorentz force, $\mathbf{j} \times \mathbf{B}$, points towards the vacuum in either expansions or contraction phases, thus making the interface always unstable to the MRTI. The simulation in Fig. \ref{figInstability}a shows a zoom of the shocked plasma layer where perturbations first appear. The magnitude of the effective acceleration $g_{\text{eff}}$ can be inferred from the simulations by tracking the position of the edge of the cavity in time. A simple but accurate estimate can be obtained by balancing the Lorentz and the ram pressure forces at stagnation (maximum radius of expansion),   $g_{\text{eff}}\sim j B \rho^{-1}\sim v_{\bot}^{2} \delta^{-1}_{sl}\sim 5\times 10^{13}$~m~s$^{-2}$, with $v_{\perp}\sim 100$~km s$^{-1}$.

The observation of dominant flute-like modes is coherent with anisotropy introduced by MRTI on the growth rates for modes parallel and perpendicular to it. In the incompressible limit, which is valid here given that the Atwood number is equal to one \cite{baker1983}, the growth rate for a mode with a wavevector $\mathbf{k}$ is given by\cite{chandrasekhar1961hydrodynamic} $\gamma^{2}=kg_{\text{eff}} - 2(\mathbf{k} \cdot \mathbf{B})^{2}/(\mu_{0}\rho)$.  The fastest growing modes are interchange modes ($\mathbf{k} \cdot \mathbf{B}= 0$), while modes with a finite value of $\mathbf{k} \cdot \mathbf{B}$ have growth rates that can be drastically reduced  by magnetic tension. Given that interchange modes have $\gamma \propto k^{1/2}$, the experimental observation of a well defined, dominant wavenumber indicates the presence of damping mechanisms\cite{hammer1996,RyutovZpinch2000} which tend to stabilize the larger wavenumbers\cite{0741-3335-53-9-093001}. For our plasma conditions we consider the effects of finite resistivity and viscosity.  Finite ion Larmor radius effects may also reduce the growth of short-wavelengths perturbations\cite{huba1996} but in our case the ions are generally too collisional and unmagnetized for these effects to be important, the ion Hall parameter being $\omega_{ci}\tau_i\sim 0.06$. The contribution of magnetic diffusivity, $D_M$, and kinematic viscosity, $\nu$, to the dispersion relation\cite{0029-5515-1-1-003} gives a growth rate for interchange modes: $\gamma \sim (g_{\text{eff}}k)^{1/2}-k^{2}(\nu+D_{M})$. Using the Spitzer-H{\"a}rm expression for the resistivity and Braginskii's expression for the ion dynamic viscosity, and maximizing the growth rate for fixed plasma parameters, we can then find the wavelength of the fastest growing mode expected to be observed
\begin{equation}
 \lambda_{\max}[mm]\approx \pi g_{\text{eff}}^{-1/3} \left(\frac{\sqrt{A}T^{5/2}}{\Lambda \left\langle Z \right\rangle^{4}\rho} + 7.6\times10^{7}\frac{\Lambda \left\langle Z \right\rangle}{T^{3/2}}\right)^{2/3}
 \label{lambdaMax}
\end{equation}
where $g_{\text{eff}}$ and $\rho$ are in SI units and the temperature is in energy units (eV),  $\Lambda$ is the Coulomb logarithm, $A$ is the atomic number, and $T=T_i \sim T_e$. The dependence on temperature of $\lambda_{\max}$, as well as its corresponding growth time  $\gamma^{-1}_{\max}$, for a fully ionized Carbon plasma, are shown in Fig. \ref{figInstability}b. As the temperature increases, resistive damping becomes less efficient and the most unstable wavelength, over the temperature range $\sim 200-800$~eV, flattens to a narrow band of values $\lambda_{max}\sim 0.5-1$~mm. For these modes the e-folding time is less than two nanoseconds. Calculation for Teflon over this range of temperature, and for $\left\langle Z \right\rangle\sim 6-8$, gives similar results. The predicted growth time-scale of $\lesssim 2$~ns is consistent with the appearance of flutes as early as 8 ns, and as shown in Fig.\ref{fig_experiment}f, the initial growth of the instability (i.e. before the collapse of the cavity which takes place around 20 ns) leads to density modulations with $\lambda \lesssim 1$~mm, in good agreement with the simulations. The subsequent increase in time of seen in Fig. \ref{fig_experiment}f can be attributed to the stretching of the slab whose velocity increases along the $z$-direction ($v \propto z/t$\cite{Higginson2017}). Finally, the plasma temperature expected ($\gtrsim 200$~eV) are consistent with the simulations and with X-ray spectroscopy data, whose analysis provides a lower-limit estimate on the electron temperature in the shocked part of the plasma of $T_e\sim 240$~eV (see Supplemental Material).

In conclusion, our results demonstrate that the expansion of a plasma away from a laser irradiated target is not easily hampered by a transverse magnetic field. Instead, the plasma dynamics is more complex and intrinsically three-dimensional, with the magnetic field confining the plasma into a rapidly expanding slab via a series of re-collimating shocks. Even for modest laser irradiations ($I \sim 10^{13}$ W cm$^{-2}$) and relatively strong fields (20 T), the plasma expands to distances over 20 times the laser focal spot diameter. We have highlighted the development of the fluid-like, MRTI. The strong magnetic field plays a critical role in driving the instability through an effective acceleration in the rest frame of the interface separating the plasma and magnetized vacuum, as well as stabilizing modes with finite values of $\mathbf{k} \cdot \mathbf{B}$\, leaving thus only interchange modes ($\mathbf{k} \cdot \mathbf{B}$\ =0) to grow. Our platform opens the door to studies of the MRTI instability in laser-produced plasmas. These could expands the research done on z-pinches \cite{lebedev_1998} to different regimes (uniform and easily controllable magnetic field strengths, higher effective accelerations, etc.) and geometries (non-axisymmetric). Nevertheless, how far the instability can be driven into the non-linear regime remains to be investigated. In addition, we note that our platform could be used to study the physics of the propagation of waves in magnetically-structured inhomogeneous mediums, such as those encountered in the solar atmosphere (e.g. \cite{1978ApJ...226..650I,1979ApJ...227..319W,Arregui2015}). Indeed, at later times the elongated slab develops kink-like modes that affect the whole body of the plasma ( $z >10$~mm in Fig. \ref{fig_experiment}c). As there is no acceleration, these modes are not due to the MRTI, but we suggest that they may be driven by the non-zero bulk velocity of the plasma in $z$-direction which destabilizes the magneto-acoustic normal modes propagating within the slab\cite{Roberts1981,Edwin1982,1992ApJ...399..478H}.

\section*{Acknowledgements}
The authors acknowledge the support of the LULI teams for technical support, the Dresden High Magnetic Field Laboratory at Helmholtz-Zentrum Dresden-Rossendorf for the development of the pulsed power generator, B. Albertazzi and M. Nakatsutsumi for their previous work in laying the groundwork for the experimental platform. This work was supported by funding from the European Research Council (ERC) under the European Union’s Horizon 2020 research and innovation program (Grant Agreement No. 787539). This work was partly done within the Plas@Par LABEX project and supported by grant 11-IDEX-0004-02 from ANR (France). The IAP RAS members acknowledge the support of the Russian Foundation for Basic Research (RFBR) foundation in the frame of projects \#18-29-21018 and \#18-02-00850. X-ray data measurement, modeling and analysis is made by the JIHT RAS (Joint Institute for High Temperatures Russian Academy of Sciences) team under financial support of Russian Science Foundation (project \#17-72-20272). O.W. would like to acknowledge the German Research Foundation Programmes GRK 1203 and SFB/TR18. Part of the experimental system is covered by a patent [no. 1000183285, 2013, Institut National de la propri\'et\'e industrielle (INPI), France]. Part of this work was supported by the region Ile-de-France through the program DIM ACAV. The research leading to these results is supported by Extreme Light Infrastructure Nuclear Physics (ELI-NP) Phase I, a project co-financed by the Romanian Government and European Union through the European Regional Development Fund. This work was granted access to the HPC resources of MesoPSL financed by the Region Ile de France and the project Equip\@Meso (reference
ANR-10-EQPX-29-01) of the programme Investissements d’Avenir supervised by the Agence Nationale pour la Recherche

\bibliography{biblio.bib}

\end{document}